\documentstyle[twocolumn,aps,epsf]{revtex}
\draft 

\begin{document}

\noindent {\bf Comment on "Swarming Ring Patterns in Bacterial Colonies Exposed
to Ultraviolet Radiation"}        

\bigskip
In an interesting experiment, Delprato {\it et
al.}~\cite{Delprato/Samadani/Kudrolli/Tsimring:2001} observed pattern formation
by colonies of soil bacteria {\it Bacillus subtilis} exposed to ultra violet
(UV) radiation on a 2-dimensional surface. Under spatially uniform UV radiation
exposure (30 W/m$^2$) the population, initially uniformly distributed over a
nutrient rich agar surface, was observed to move toward the edge of the colony
until, after approximately 45 hours, virtually the entire population was
concentrated within a few mm of  the colony's edge, forming a ring. After
removal of the UV, the colony grew normally outward and inward from the ring.
Thus the authors ruled out nutrient depletion near the colony center as a
reason for evacuation~\cite{Budrene/Berg:1991}. The authors offer a possible
explanation based on a variant of the usual reaction-diffusion
dynamics~\cite{Tsimring:1995}, whereby the bacteria, stimulated by the UV,
migrate towards the edge of the colony because of reduced waste concentration
there.

We point out here that the formation of a ring pattern is an artifact of the
confinement of the dynamics to two dimensions. As we show by example below, the
only conditions necessary for the formation of ring patterns is that the
agents must flee UV, move in a 2-dimensional geometry, and be confined
to a circular area.

{\it Bacillus subtilis}, a motile bacteria, normally lives and moves in a
3-dimensional environment.  A variety of living creatures flee UV for one
reason or another and when confined to two dimensions make ring patterns. One
example is the well investigated fresh water zooplankton {\it Daphnia magna}.
These plankton are a favorite prey of fish that hunt visually. For a variety of
reasons including temperature and food  {\it Daphnia} stay near the water
surface at night. With the rising of the sun, UV in the surface layer of the
water signals {\it Daphnia} to swim further down the water column where
darkness  and mud concentration shield them from most predatory fish and also
from harmful UV radiation~\cite{Rhode:2001} (this process is commonly known as
diel vertical migration).

Figure 1(a),(b) shows  a {\it Daphnia} population, approximately uniformly
distributed in dark, immediately  fleeing to the bottom of a 3-dimensional
medium (15 cm of water in a $50 \times 20$ cm aquarium) upon exposure of the top
surface to UV. When confined to an approximately 2-dimensional medium (1 cm of
water in a 15 cm diameter Petri dish) and exposed to a uniform intensity of UV
{\it Daphnia} immediately flee to the edge of the dish (Fig. 1(c),(d))
thus forming a ring similar to that observed by Delprato {\it et al.} These
behaviors are robust in both two and three dimensions having been observed many
times under various conditions.

In contrast to the reasons offered above by Delprato {et al.}, we propose that
{\it Bacillus subtilis} flee UV in order to avoid the sun drying out the surface
layers of soil in a manner related to the classic motions of the magnetotactic
bacteria, {\it Spirochaeta plicatilis}~\cite{Blakemore:1975} to which they are
similar.  Therefore the above given conditions necessary for the formation of
ring patterns are satisfied in the experiment of Delprato {\it et al.}: The
agents flee UV, are forced to move in a 2-dimensional geometry, and are confined
to a circular area given by the shape of the colony grown before the UV radiation
is switched on. Thus the experimental design of Delprato {\it et al.} is
artificial as neither bacteria nor {\it Daphnia} do naturally live in a
2-dimensional environment. To investigate the reaction of {\it Bacillus subtilis}
to UV radiation coming from above, a seminal experiment should have a
3-dimensional setup.

\begin{figure}
\begin{center}
\unitlength 0.3mm
\def\epsfsize#1#2{0.4#1}
\begin{picture}(285,258)
\def\epsfsize#1#2{0.163#1}
\put(14,0){\epsfbox{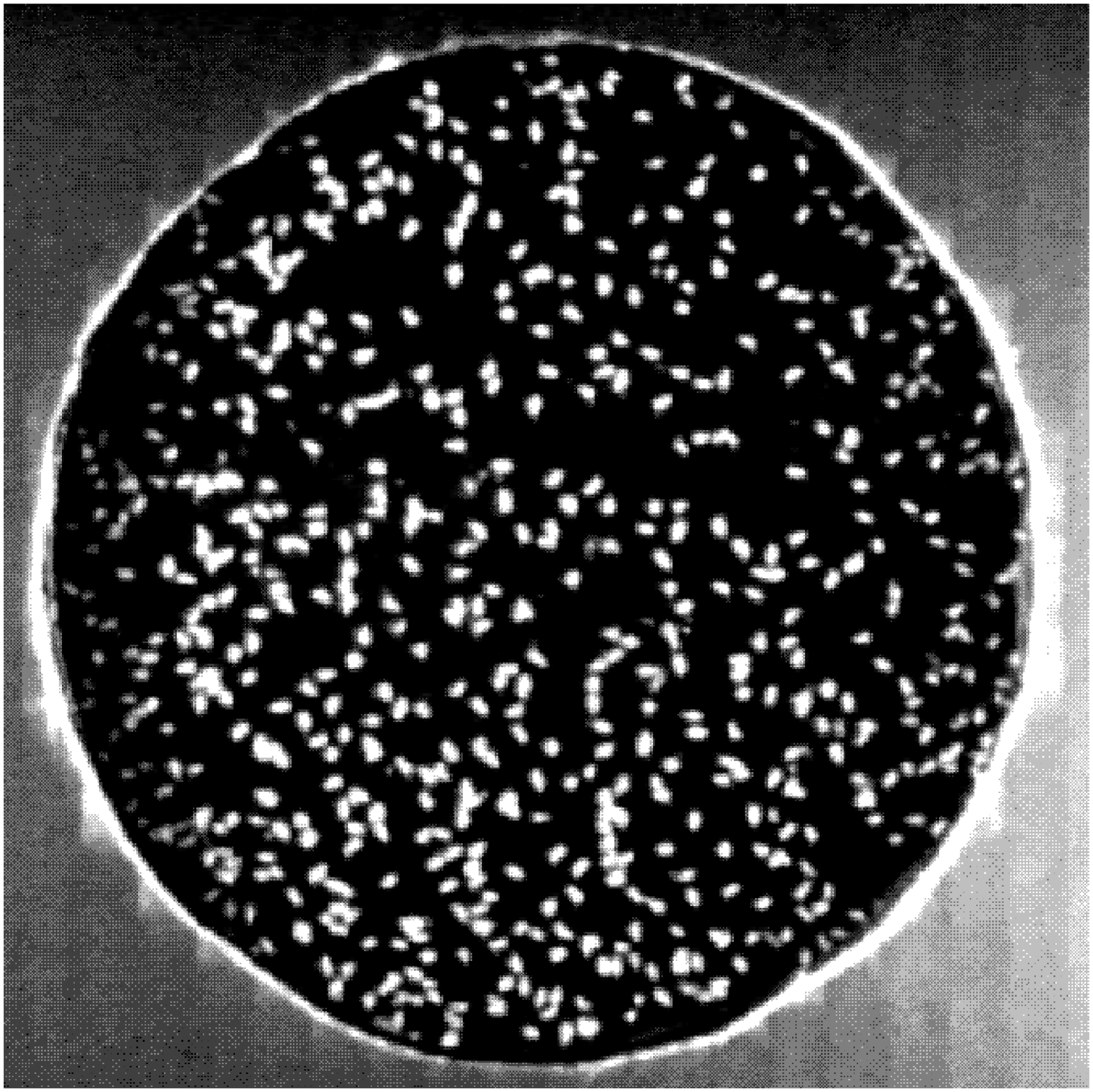}}
\put(145,0){\epsfbox{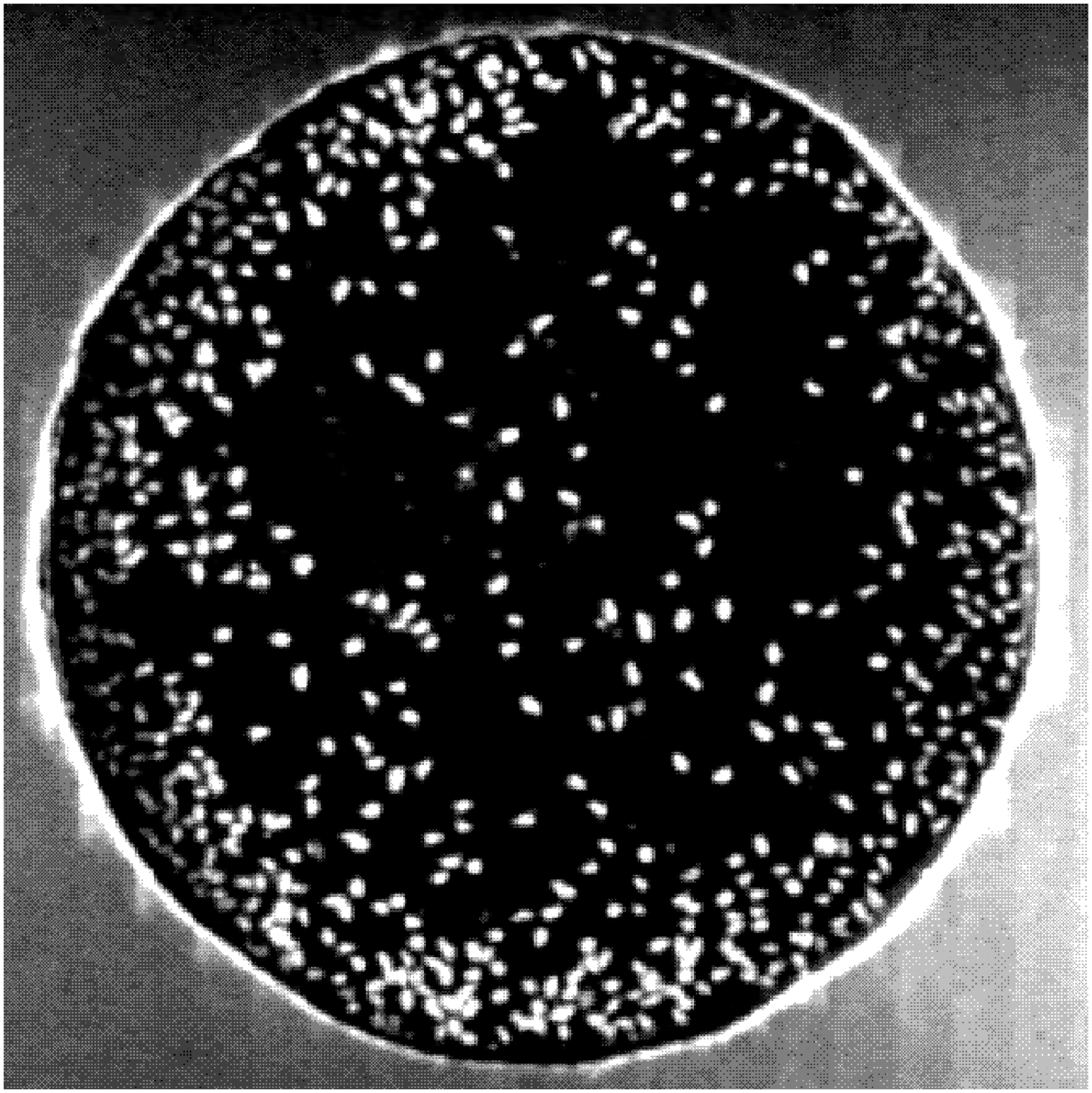}}
\put(145,132){\epsfbox{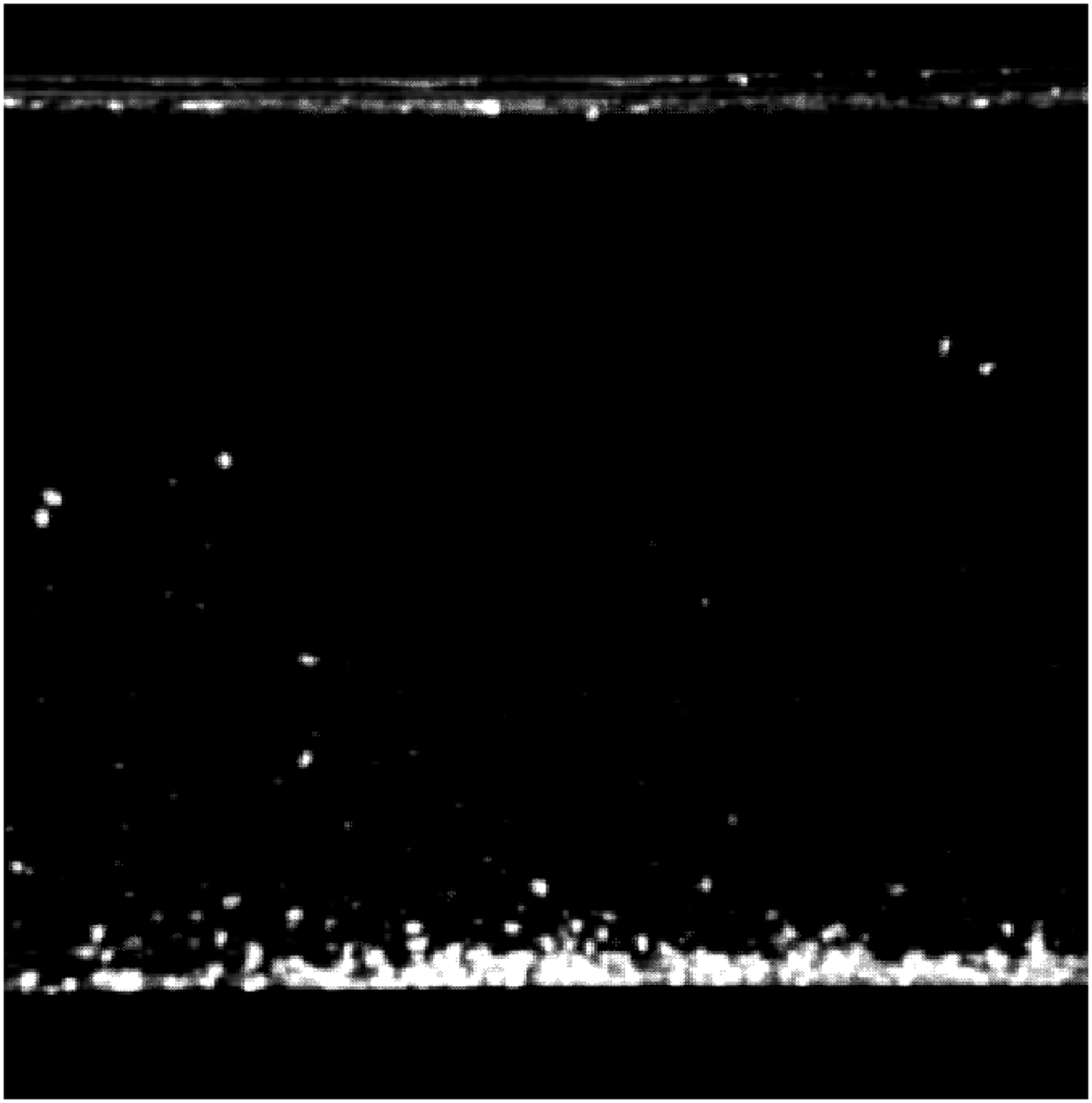}}
\put(14,132){\epsfbox{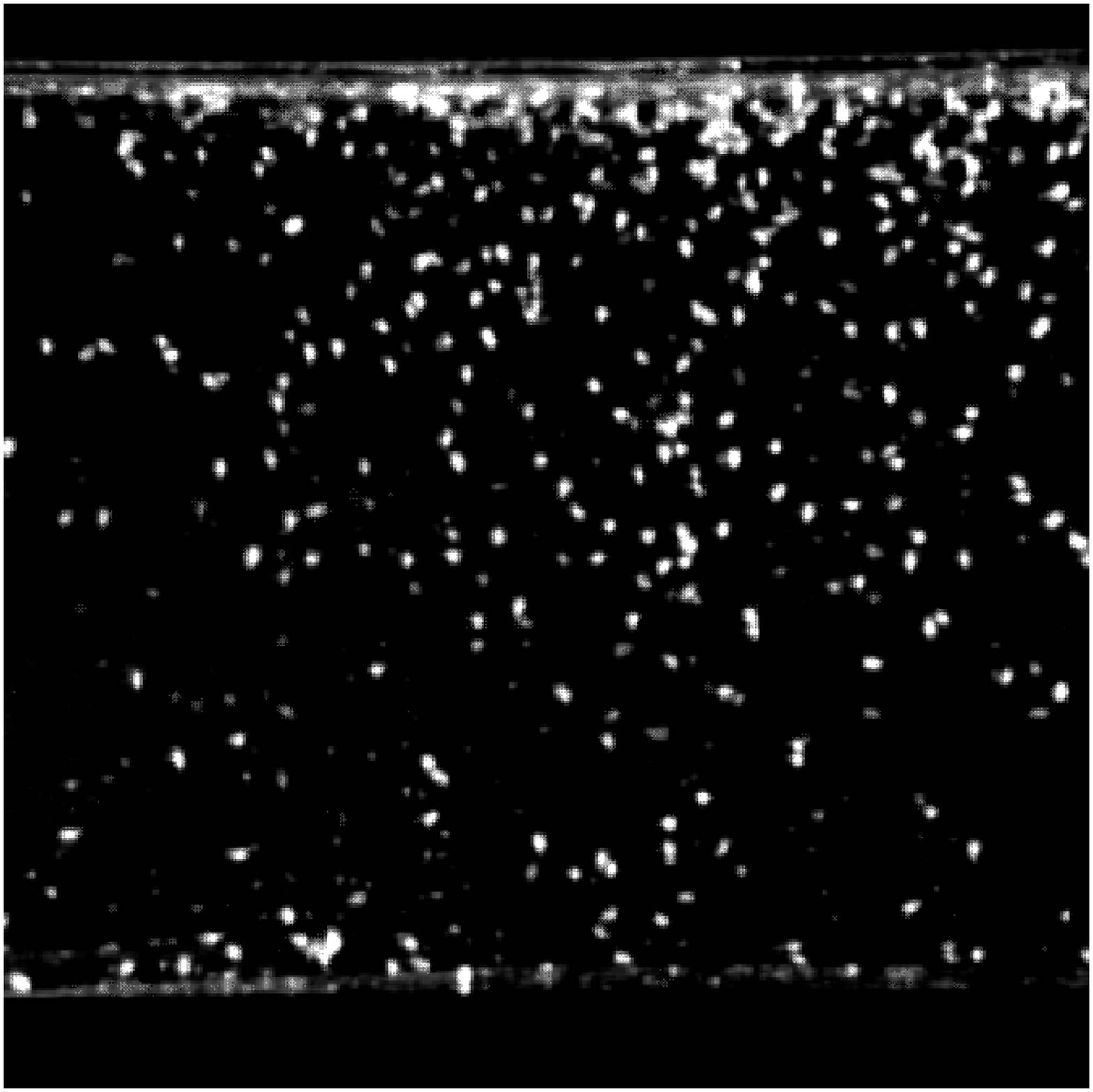}}
\def\epsfsize#1#2{0.145#1}
\put(14,244){\epsfbox{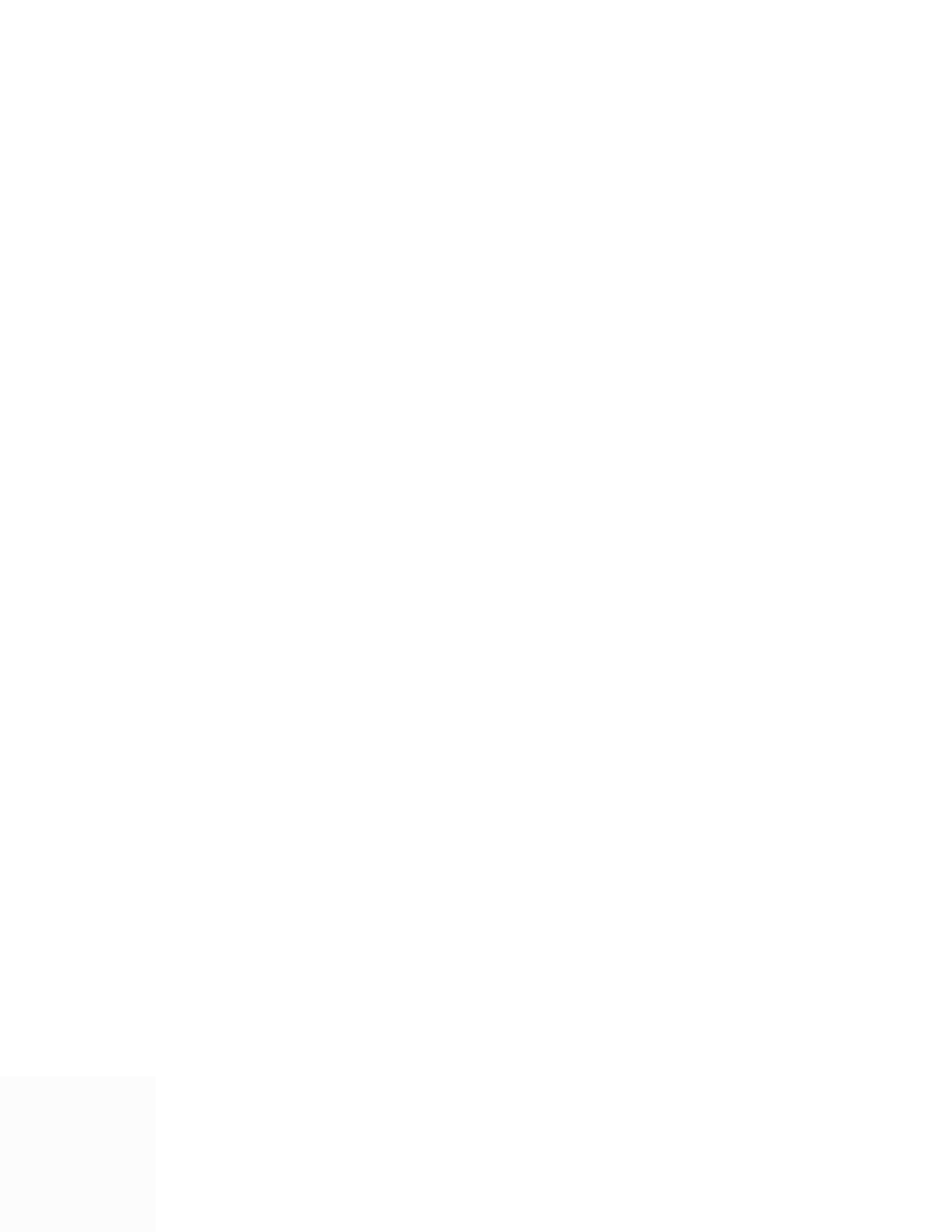}}
\put(145,245){\epsfbox{whitesquare.eps}}
\put(14,112){\epsfbox{whitesquare.eps}}
\put(145,112){\epsfbox{whitesquare.eps}}
\def\epsfsize#1#2{0.4#1}
\put(17,252){\makebox(10,5)[t]{(a)}}
\put(148,254){\makebox(10,5)[t]{(b)}}
\put(17,121){\makebox(10,5)[t]{(c)}}
\put(148,121){\makebox(10,5)[t]{(d)}}
\end{picture} 
\end{center}
\caption{Pictures of {\it Daphnia} (illuminated by infrared to which they are
blind) in the dark and exposed to UV radiation from top (Blak-Ray UV lamp,
radiating at 365nm, 20 W/m$^2$ at the water surface).
(a) Sideview of the aquarium in the dark, (b)  after 25
seconds of UV exposure. (c) View from below on Petri dish in the dark, (d)
after 30 seconds of UV exposure ({\it Daphnia} visible in the center region are
mostly motionless, shocked by the hazardous UV light).} \label{fig:Daphnia}
\end{figure}

\bigskip

\noindent A.O. gratefully acknowledges the Alexander von Humboldt Foundation 
for financial support.

\bigskip
\noindent Anke Ordemann and Frank Moss\\
Center for Neurodynamics,\\
University of Missouri at St. Louis,\\
St. Louis, MO 63121

\bigskip

\noindent Received June 8, 2002 

\end{document}